\documentclass{iaus}
\usepackage{graphicx}

\title[Searching for transit timing variations] %% give here short title %%
{Searching for transit timing variations in transiting exoplanet systems}

\author[Marie Hrudkov{\'a}]   %% give here short author list %%
{Marie~Hrudkov{\'a}$^1$, Ian~Skillen$^2$, Chris~Benn$^2$, Don~Pollacco$^3$,
Neale~Gibson$^3$, Yogesh~Joshi$^3$, Petr~Harmanec$^1$ and Simon~Tulloch$^2$} 

\affiliation{$^1$Astronomical Institute of the Charles University, \\
Faculty of Mathematics and Physics,
V Hole\v{s}ovi\v{c}k{\'a}ch 2, 18000, Praha 8, Czech Republic \\
email: {\tt mariehrudkova@gmail.com}\\[\affilskip]
$^2$Isaac Newton Group of Telescopes, \\
Apartado de Correos 321, E-38700, Santa Cruz de la Palma, Canary Islands, Spain\\[\affilskip]
$^3$Queen's University Belfast, \\
University Road, BT7 1NN, Belfast, UK} 

\pubyear{2008}
\volume{253}  %% insert here IAU Symposium No.
\pagerange{1--4}
\jname{Transiting Planets}
\editors{A.C. Editor, B.D. Editor \& C.E. Editor, eds.}
\begin{document}

\newcommand{\jed}[1]{~{\rm #1}}

\hyphenation{si-mu-la-ti-ons ack-now-led-ge sup-port Re-se-arch Pro-gram}

\maketitle

\begin{abstract}
Searching for transit timing variations in the known transiting
exoplanet systems can reveal the presence of
other bodies in the system. Here we report such searches for two transiting
exoplanet systems, TrES-1 and WASP-2. Their new transits were observed with the 4.2m
William Herschel Telescope located on La Palma, Spain. In a continuing
programme, three consecutive transits 
were observed for TrES-1, and one for WASP-2 during September 2007. We used
the Markov Chain Monte Carlo simulations to derive transit times and
their uncertainties. The resulting transit times are consistent with the
most recent ephemerides and no conclusive proof of additional bodies in either system
was found.
\end{abstract}

\firstsection % if your document starts with a section,
\section{Introduction}

Transiting planets provide a wealth of information about exoplanetary
systems. Short-term variations in the mid-eclipse times of the transits may 
reveal the presence of moons, trojans or other
planets (\cite[Holman \& Murray 2005]{HolmanMurray2005}, \cite[Agol \etal\
2005]{Agol_etal05}, \cite[Ford \& Holman 2007]{FordHolman07}), whereas long-term variations 
could result from orbital precession (\cite[Miralda--Escud\'{e}
2002]{Miralda02}).
This provides further constraints on theories of
planetary system formation and evolution, as well on theories of planetary
atmospheres and their composition. 

\section{Observations and Data Reduction}

We observed three transits of TrES-1 on UT 2007 September 12, 15, and 18,
corresponding to epochs $E=152$, 153, and 154 of the ephemeris given by
\cite[Winn \etal\ (2007)]{Winn_etal07}:
\begin{equation}
T_c(E)={\rm HJD}\,(2453895.84297\pm 0.00018)+(3^{\rm d}\!\!.0300737\pm
0^{\rm d}\!\!.0000026)\times E,\label{efe-tres}
\end{equation}

\noindent
and one transit of WASP-2 on UT 2007 September 13, corresponding to epoch
$E=162$ of the ephemeris given by \cite[Charbonneau \etal\
(2007)]{Charbonneau_etal07}:
\begin{equation}
T_c(E)={\rm HJD}\,(2454008.73205\pm 0.00028)+(2^{\rm d}\!\!.152226\pm
0^{\rm d}\!\!.000004)\times E.\label{efe-wasp}
\end{equation}

We used AG2, a frame-transfer CCD mounted at the folded Cassegrain focus of
the William Herschel Telescope of the Isaac Newton Group, La Palma, Spain. AG2 is
an ING-designed autoguider head with E2V CCD having a field of view (FOV) 
$3.3\times 3.3\jed{arcmin^2}$ and a
pixel scale of $0.4\jed{arcsec/pixel}$. We used a Kitt Peak R filter in order
to minimize the effect of color-dependent atmospheric extinction on the
differential photometry and the effect of limb-darkening on the
transit light curve. 

We observed under nearly perfect conditions on the nights of UT September
12, 13, and 18. On the night of UT September 15, we observed under scattered clouds in the
second half of the night. On all nights, autoguiding ensured the positions of
all stars on the CCD varied by no more than 3 pixels over the course of each
night. We strongly defocused the telescope in order to minimize the effect of the
pixel-to-pixel sensitivity variations, and also to enhance the duty cycle.
The full-width at half-maximum (FWHM) of stellar images was 7 pixels for
WASP-2 and ranged from 8 to 13 pixels for TrES-1. For TrES-1 for the three
nights, we acquired 5, 10 and 10 second exposures, while for WASP-2 we used 
7 second exposures.    

We used standard IRAF\footnote{The Image Reduction and Analysis Facility
(IRAF) is distributed by the National Optical Astronomy Observatories, which
are operated by the Association of Universities for Research in Astronomy,
Inc., under cooperative agreement with the National Science Foundation.} procedures
for the overscan correction, bias subtraction and performing the differential
photometry. We did not use flat-fields to minimise the impact of
position-angle dependent scattered light entering the aperture. Different
aperture sizes were tried in order to find out the one that produced the minimum
noise in the out-of-transit data. For TrES-1 for the three nights, the
optimum aperture radii were 31, 31, and 34 pixels, while for WASP-2 it was 30
pixels. The sky background was subtracted, using an estimate of its
brightness determined within an annulus centered on each star.
Because of the strong defocusing and other field stars, we set an inner radius
of the annulus equal to the aperture radius and used the width of 10 pixels.
%for TrES-1 and 5 pixels for WASP-2. 
The comparison stars used were 2MASS: J19041058+3638409, J19040934+3639195, 
and J19040792+3640116 for TrES-1, and 2MASS: J20304846+0627355, and J20305168+0628008
for WASP-2.
For each night, differential photometry was obtained by taking
the ratio of the signal of the variable to the mean of the comparisons, both
normalized by a smooth function of time. 

\begin{figure}
\centering
\rotatebox{270}{\resizebox{5cm}{!}{\includegraphics{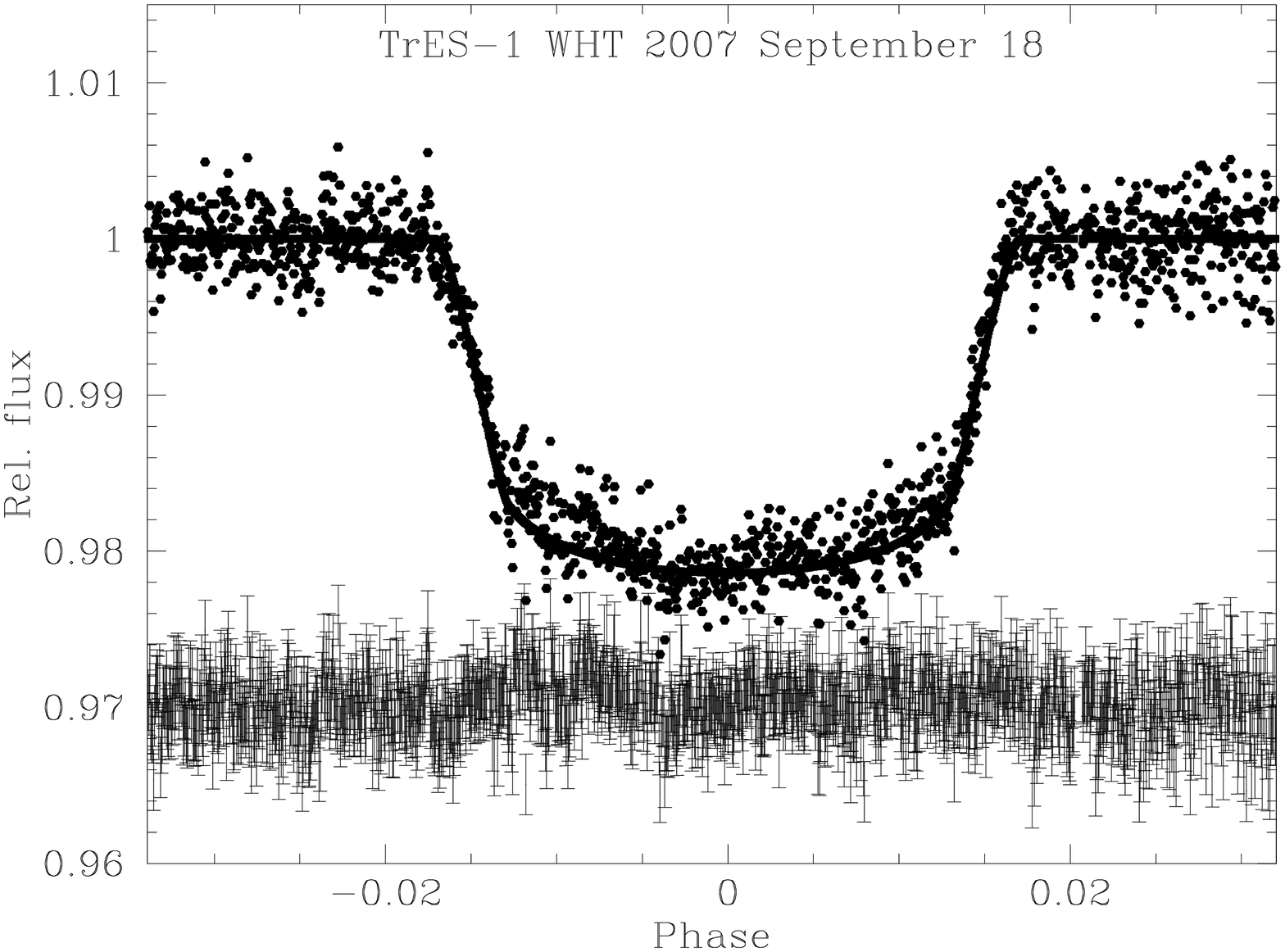}}}
\rotatebox{270}{\resizebox{5cm}{!}{\includegraphics{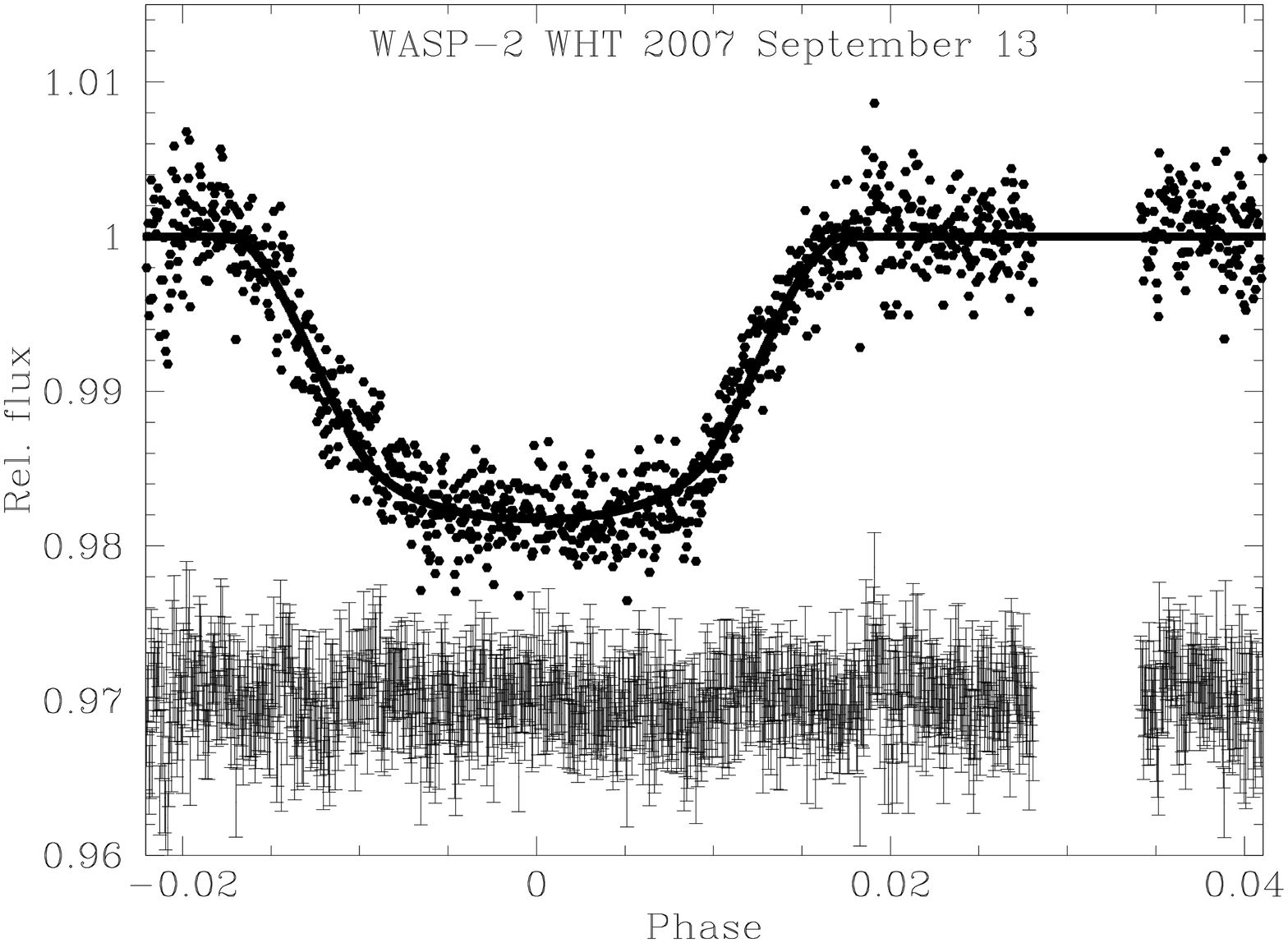}}}
\caption[]{Relative photometry for TrES-1 (left) and WASP-2 (right) is shown with the best-fitting
model (solid line). The residuals and
rescaled 1~$\sigma$ error bars are also protted, offset by
a constant flux for clarity.} \label{wt_lc}
\end{figure}

To estimate appropriate error bars for our data, we used a procedure similar
to that by \cite[Narita \etal\ (2007)]{Narita_etal07}. We first fitted the light curves with the analytic
formulae of \cite[Mandel \& Agol (2002)]{MandelAgol02} to find the
differences between the data and the best-fitting model. We rescaled
the error bars to satisfy $\chi^2/N_{DOF}=1.0$, where $N_{DOF}$ is the
number of the measurements in each light curve. For the three nights of TrES-1
we found that the true errors are higher by factors of 2.2, 1.6 and 1.9
respectively than the errors including only the Poisson noise, and for
the one night of WASP-2 by a factor of 2.2. Differential photometry with
these rescaled error bars and the best-fitting model (solid line) are shown in Fig.
\ref{wt_lc}.

In order to estimate the amount of
the time-correlated red noise, we solved the equations given e.g. by \cite[Narita
\etal\ (2007)]{Narita_etal07}:
\begin{equation}
\sigma_1^2=\sigma_w^2+\sigma_r^2,
\end{equation}
\begin{equation}
\sigma_N^2=\frac{\sigma_w^2}{N}+\sigma_r^2,
\end{equation}

where $\sigma_1$ is the standard deviation of each residual and $\sigma_N$
is the standard deviation of the average of the successive $N$ points. We
selected $N$ to correspond to about 20-50 minutes depending on the length of
the interval with the correlated errors. $\sigma_w$ is the white noise,
which is uncorrelated and averages down as $(1/N)^{1/2}$, $\sigma_r$ is the
red noise, which is correlated and remains constant for specified $N$. We
adjusted the error bars by multiplying $[1+N(\sigma_r/\sigma_w)^2]^{1/2}$
and used these rescaled uncertainties for the subsequent fitting procedure.

\begin{table}   
  \begin{center}
  \caption{Resulting barycentric transit times of TrES-1 and WASP-2 systems
with their uncertainties given by the 68\% confidence limits.}
  \label{times}
  {\scriptsize
  \begin{tabular}{lccc}\hline
{\bf System} & {\bf Epoch} & {\bf Mid-transit time} & {\bf Uncertainty}\\
\hline TrES-1&152&{\rm BJD} 2454356.41417&0.00010\\
TrES-1&153&{\rm BJD} 2454359.44431&0.00015\\
TrES-1&154&{\rm BJD} 2454362.47424&0.00020\\
WASP-2&162&{\rm BJD} 2454357.39254&0.00016\\ \hline
  \end{tabular}
  }
 \end{center}  
\end{table}

\section{The Model}

We used the Markov Chain Monte Carlo simulations with the
Metropolis-Hastings algorithm (\cite[Ford 2005]{Ford05}) to estimate the
statistical uncertainties of the resulting parameters. We assumed a
quadratic law of the limb darkening. For TrES-1, we fixed the system parameters
according to \cite[Winn \etal\ (2007)]{Winn_etal07} and solved only for the transit times. For
WASP-2, we fixed the parameters derived by \cite[Charbonneau \etal\
(2007)]{Charbonneau_etal07} and
fitted the limb-darkening coefficients and the transit time. For each
night, we created 10 independent chains, with the length typically
100,000 points in each chain. We checked the convergence of generated chains
using the \cite[Gelman \& Rubin (1992)]{GelmanRubin92} R statistic.  

\section{Results}

\begin{figure}
\centering
\rotatebox{270}{\resizebox{4.6cm}{!}{\includegraphics{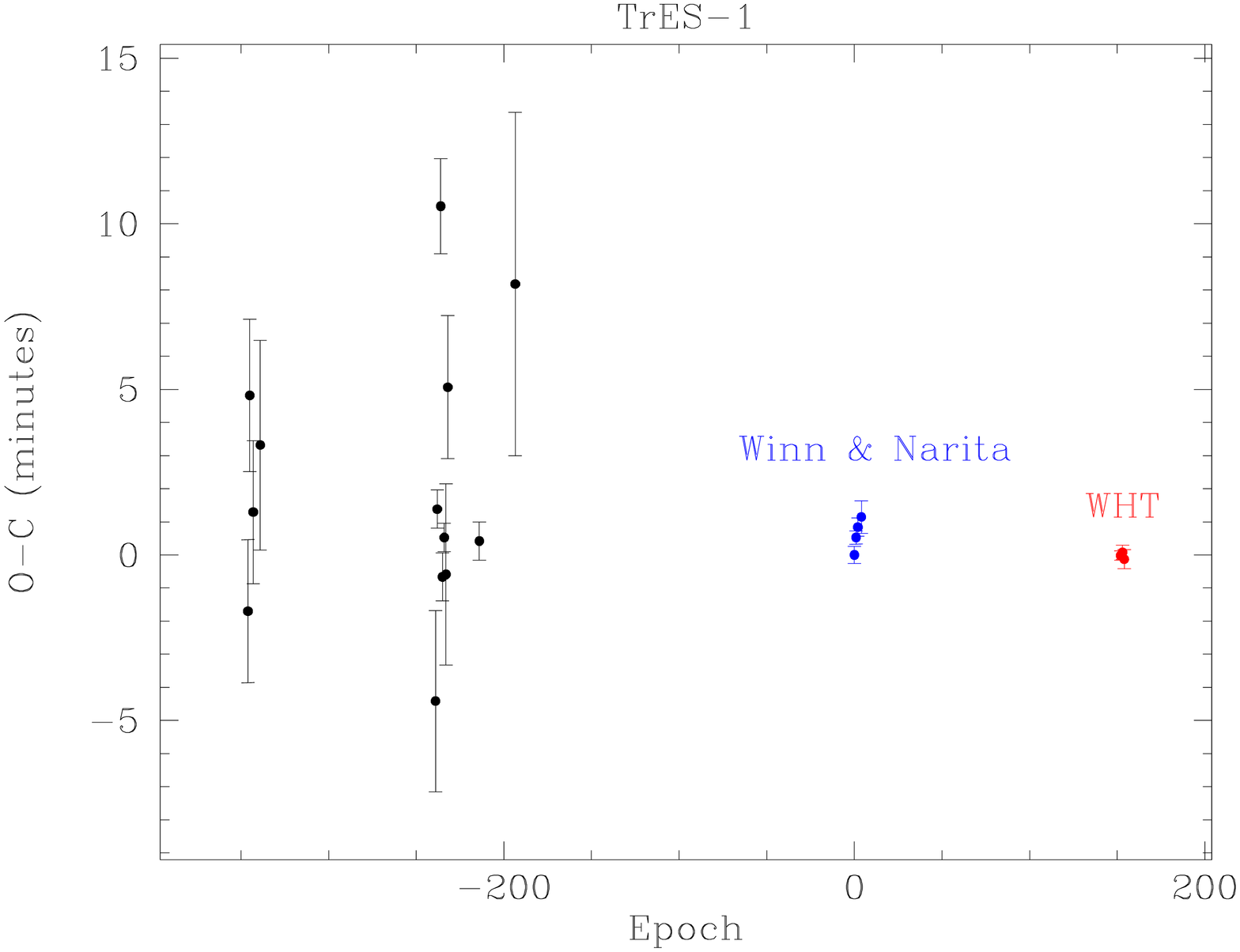} }}
\rotatebox{270}{\resizebox{4.6cm}{!}{\includegraphics{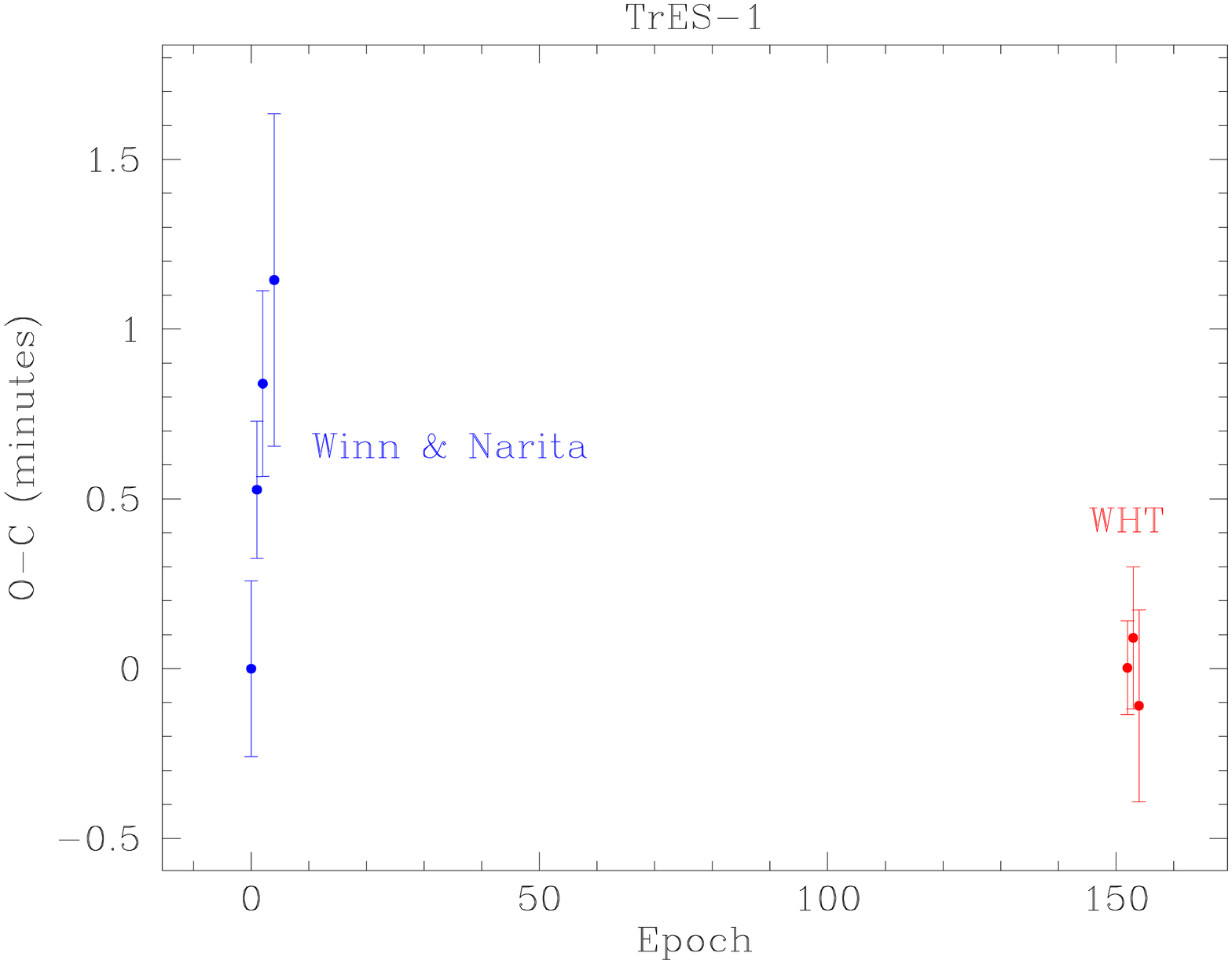} }}
\caption[]{\hbox{$O\!-\!C$} residuals of mid-transit times of TrES-1 system. 
{\bf Left.} \hbox{$O\!-\!C$} of TrES-1 including the data from \cite[Charbonneau \etal\ (2005)]{Charbonneau_etal05},
(their Table 1), \cite[Winn \etal\ (2007)]{Winn_etal07}, \cite[Narita \etal\
(2007)]{Narita_etal07} and our three nights from WHT. They were computed using the ephemeris of \cite[Winn \etal\
(2007)]{Winn_etal07}. {\bf Right.} \hbox{$O\!-\!C$} in detail indicating no significant transit timing variations.} 
\label{tres_oc}
\end{figure}

\begin{figure}
\centering
\rotatebox{270}{\resizebox{4.6cm}{!}{\includegraphics{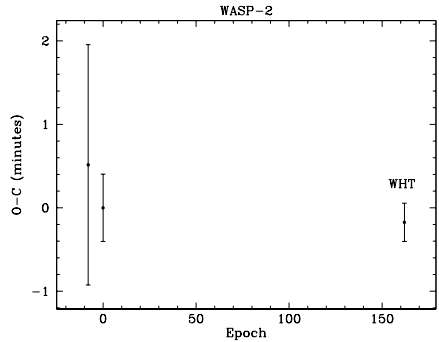} }}
\caption[]{\hbox{$O\!-\!C$} residuals of mid-transit times of WASP-2 system show the resulting transit time is
consistent with the ephemeris given in \cite[Charbonneau \etal\
(2007)]{Charbonneau_etal07}.} \label{wasp_oc}
\end{figure}

The resulting barycentric transit times can be found in Table \ref{times}.
Their uncertainties are given by the 68\% confidence limits.
The resulting \hbox{$O\!-\!C$} residuals are plotted in Fig.~\ref{tres_oc}
for TrES-1, and in Fig.~\ref{wasp_oc} for WASP-2, using the
ephemerides (\ref{efe-tres}), and (\ref{efe-wasp}), respectively.
In summary, our data provide no conclusive evidence of other bodies in either
system but the systematic trend of \hbox{$O\!-\!C$} residuals for TrES-1,
found by \cite[Winn \etal\ (2007)]{Winn_etal07}, is worthy of further study.
For TrES-1, our 3 new transit times plus 1 by
\cite[Narita \etal\ (2007)]{Narita_etal07} extend the interval covered
by data for more than 150 cycles. We therefore used a weighted
linear fit to the transit times as a function of epoch (weights
inversely proportional to the squares of rms errors of individual transit
times) to derive a more accurate ephemeris for the future transit
predictions:
\begin{equation}
T_c(E)={\rm HJD}\,(2453186.80625\pm 0.00054)+(3^{\rm d}\!\!.0300725\pm
0^{\rm d}\!\!.0000027)\times E.\label{new-tres}
\end{equation}\\

\begin{acknowledgements}{}

The research was supported by the grants
205/08/H005 and 205/06/0304 of the Czech Science Foundation and also from
the Research Program MSM0021620860
of the Ministry of Education of the Czech Republic.
%We acknowledge the use of the electronic
%bibliography maintained by NASA/ADS system and by the CDS in Strasbourg.

\end{acknowledgements}

\end{document}